\documentclass[reprint,
superscriptaddress,
 amsmath,amssymb,
 aps,
 pra, 
longbibliography
]{revtex4-2}
\usepackage[utf8]{inputenc}
\usepackage{amsmath}
\usepackage{hyperref}
\hypersetup{
     colorlinks=true,
     linkcolor=blue,
     filecolor=blue,
     citecolor = blue ,      
     urlcolor=cyan,
     }
\usepackage{braket}
\usepackage{graphicx}
\usepackage{amsfonts}
\usepackage{amssymb}
\usepackage{xcolor}

\usepackage{comment}
\usepackage{mathtools}

\begin{document}

\title{Linear Response Study of Collisionless Spin Drag}
\author{Donato Romito}
\affiliation{University of Southampton}
\affiliation{INO-CNR BEC Trento}
\author{Carlos Lobo}
\affiliation{University of Southampton}
\author{Alessio Recati}
\affiliation{INO-CNR BEC Trento}
\affiliation{University of Trento}
\date{August 2019}

\begin{abstract}
In this work we are concerned with the understanding of the collisionless drag or entrainment between two superfluids, also called Andreev-Bashkin effect, in terms of current response functions.
The drag density is shown to be proportional to the cross transverse current-current response function, playing the role of a \textsl{normal} component for the single species superfluid density.
We can in this way link the existence of finite entrainment with the exhaustion of the energy-weighted sum rule in the spin channel.
The formalism is then used to reproduce some known results for a weakly interacting Bose-Bose mixture.
We include the drag effect to determine the beyond mean-field correction on the speed of sound and on the spin dipole excitations for a homogeneous and a trapped weakly interacting gas, respectively.
Finally we show that the response  to a quick dipole perturbation on one of the species induces a dipole moment on the other species which is proportional to the drag at short times.
\end{abstract}

\maketitle

\section{Introduction}
The superfluid drag was predicted by Andreev and Bashkin \cite{AndBas} for a two component superfluid mixture, correcting previous works by Khalatnikov \cite{Khalatnikov}.
The effect predicts that the superflow in one component will induce a superfluid current in the other one without any dissipation.
In other words the superfluid currents of each component in a mixture depend on both superfluid velocities \cite{AndBas}.
Although the Andreev-Bashkin effect was first predicted \cite{AndBas} in the context of $^3$He-$^4$He mixtures, a mixture of these two components where both are in the superfluid state cannot be achieved experimentally, due to their low miscibility.
The Andreev-Bashkin effect has also been discussed in the hydrodynamics of neutron star cores (see \cite{LatPra} and references therein), which are believed to be made of a mixture of superfluid neutrons and protons. Cold atomic mixtures have also been proposed as a promising environment \cite{FilShe,LinSub,NesRec,ParGio,DefEns,BabSel} where the effect could be observed.
Nevertheless, a direct experimental observation of the drag is still missing.

The main aim of this paper is to describe how the drag effect arises from the general microscopic many-body theory of two interacting quantum fluids as well as its effect on their dynamics. 
%We will consider a miscible mixture of %two superfluids consisting of two %types of atoms whose number is %conserved separately and which have %velocity independent interactions. 
In particular we relate the superfluid drag density to the current-current response functions, making a clear distinction between their transverse and longitudinal long wavelength limits.
This approach applies to any quantum mixture in the linear response regime. Therefore it can be employed to predict the magnitude of the drag effect in a variety of systems such as Bose-Bose superfluid mixtures on a lattice, Bose-Fermi superfluid mixtures and Fermi-Fermi superfluid mixtures.

We also connect this result with the formalism of sum rules, which is an established tool to study the elementary and collective excitations of (trapped) quantum gases (see \cite{StrPit}). We show how the presence of the drag results in a correction to the energy weighted sum rule.
We then apply the linear response formalism to the case of a weakly interacting Bose mixture with $\mathbb{Z}_2$ symmetry, from which we derive the beyond mean field frequency shifts of the elementary spin excitations.
Finally we study the linear response at short times to a perturbation that is quickly switched on, showing that it can be used to measure the drag effect in experiments.

While we will mostly focus on homogeneous systems, our formalism applies to discrete space Hamiltonians alike, with just very small modifications as explained in Appendix \ref{subsec:latticedrag}.

The structure of the paper is as follows.
In sec.~\ref{sec2} we provide the microscopic definition of the drag within linear response theory and rigorously derive its correction to the energy weighted sum rule.
In sec.~\ref{sec:WIBG} we  specialize to a $\mathbb{Z}_2$ symmetric Bose-Bose mixture.
We recover the known value \cite{FilShe} of the superfluid drag in subsec.~\ref{subsec:bogoliubov} and calculate its effect on the spin speed of sound and on the spin dipole frequency in subsecs.~\ref{sec:sound} and \ref{sec:dipole}, respectively.
In sec.~\ref{sec:quickresponse} we devise an efficient way to measure the drag by studying the short-time response to a quick perturbation.
Appendix \ref{app:thermodynamics} includes results for thermodynamic quantities such as the susceptibility and the chemical potential for a weakly interacting mixture. These results are used in subsecs.~\ref{sec:sound} and \ref{sec:dipole}.
Appendix \ref{subsec:latticedrag} is devoted to generalizing the results for Bose-Hubbard Hamiltonians.

\subsection{The Andreev-Bashkin effect: three fluid hydrodynamics}
In \cite{Khalatnikov}, Khalatnikov, inspired by the two-fluid model \cite{Landau} used in single component superfluidity, chose to describe a mixture of {\em two} components in terms of a three fluid model, i.e. the normal fluid and the two superfluids. However he did not consider the possibility of a coupling between the two superflows.
Later, Andreev and Bashkin \cite{AndBas} introduced such a coupling which gives rise to their eponymous effect. For the case of a homogeneous system in the hydrodynamic limit, to describe the relation between the mass current densities $m_\alpha \mathbf{j}_\alpha$ of each component $\alpha=A,B$ and the velocities, they introduced the matrix of the superfluid densities $\rho_{\alpha \beta}$ defined such that:
\begin{eqnarray}\label{AB}
m_A\mathbf{j}_A\!&=&\!\rho_{n A }\mathbf{v}_{n}\!\!+\rho_{A A}\mathbf{v}_{A}^{(s)}\!\!+\rho_{AB}\mathbf{v}^{(s)}_B \, , \nonumber \\
m_B\mathbf{j}_B\!&=&\!{\rho_{n B}} \mathbf{v}_{n}\!\!+\rho_{BB}\mathbf{v}_{B}^{(s)}\!\!+\rho_{BA}\mathbf{v}^{(s)}_{A}.
\end{eqnarray}
In Eq.~(\ref{AB}) $m_\alpha$ are the bare masses of the constituent atoms of component $\alpha$, $n_\alpha$ the number densities and $\mathbf{v}^{(s)}_\alpha$ are the superfluid velocities defined as $\mathbf{v}_\alpha^{(s)}=\hbar/m_\alpha \nabla \varphi(\mathbf{r},t)$, with $\varphi(\mathbf{r},t)$ the condensate phase. In the hydrodynamic regime, it is assumed that both the $A$ and $B$ normal components are in local thermal equilibrium due to their mutual collisions, implying that there is only one normal component $\rho_n=\rho_{nA}+\rho_{nB}$ moving with velocity $\mathbf{v}_{n}$.
Eq.~(\ref{AB}) describes the fact that the superflow of one component takes part in the mass current density of the other.
The off-diagonal terms of the matrix represent the drag that one component forces upon the other. The matrix $\rho_{\alpha\beta}$ is symmetric \cite{AndBas} -- as we
also show in what follows, so that $\rho_{AB}=\rho_{BA}$.

For a homogeneous system, galilean invariance requires that:
\begin{equation}\rho_{n \alpha} =m_\alpha n_\alpha-\rho_{\alpha \alpha}-\rho_{AB} \, .
\end{equation}\label{eq:normalization1}
So that at zero temperature where the normal component vanishes \cite{Leggett}, the sum of all the superfluid densities is the total mass density of the system $\rho$, namely:
\begin{equation}\label{eq:normalization2}
\rho \equiv m_A n_A +m_B n_B=\rho_{AA}+\rho_{BB}+2\rho_{AB} \, .
\end{equation}
This relation is modified in presence of a lattice since it breaks translational invariance, as shown in Appendix \ref{subsec:latticedrag}.

The Andreev-Bashkin drag is collisionless in nature: it results from the renormalization of the mass of particles of one component by the interaction with the other component.
Thus the flows of the two components are coupled without any dissipation of energy.
In this regard the collisionless drag is closely related to the polaron drag, where an impurity immersed in a bath of other indistinguishable particles (majority component) has its mass renormalized by interactions \cite{AndBas}.
This is what makes the drag non vanishing even at zero temperature.

Throughout this paper we will use the more generic term ``collisionless drag" interchangeably with ``Andreev-Bashkin drag".

\section{Microscopic description of the collisionless drag from linear response theory}\label{sec2}

In this section the three-fluid hydrodynamics of Eq.~(\ref{AB}) is connected to the microscopic theory by means of linear response.
In the context of superfluidity, linear response theory allows us to relate the superfluid and normal densities of the Landau two-fluid model with current-current response functions, as it is well described in the case of a single component (see, e.g., \cite{PinNoz,StrPit,Baym}). In subsec.~\ref{subsec:currentcurrent} we generalize this concept to the case of a two component superfluid mixture, without referring to any specific microscopic model. We shall closely follow the formalism of Refs. \cite{PinNoz,StrPit,Baym}. 
The basic idea is to start with the fluid in equilibrium and then to subject it to a weak transverse field whose intensity increases adiabatically from zero. We then can identify the current density that is imparted by the field. We will show that this is formally the same as calculating the momentum density average with a perturbed Hamiltonian where the perturbation depends on the field. This can then be expressed in terms of a current-current response function.
Finally, using Eq.~(\ref{AB}), we express the current density in terms of the $\rho_{\alpha \beta}$ and $\rho_n$, thereby relating the latter to the transverse current response function in the long wave length limit.
Although we will derive the general linear response expressions for the densities of the three-fluid model, we will mostly restrict to the zero temperature case to make the distinction with the single component case more apparent.

Aside from giving an intuitive account of the effect and making a clear connection to the existing literature on superfluidity, this approach could prove useful to make predictions on the value of the collisionless drag in various systems, thus identifying those which are the best candidates to display it in a significant way.
Another strength of this formalism is that response functions can be generally computed making use of diagrammatic theory and numerical techniques.

In the second part of this section -- subsec.~\ref{sec:sumrules} -- we will predict the effect of the drag on sum rules of the structure factor.
Sum rules provide an established method to compute the frequencies of collective oscillations for various physical systems (\cite{StrPit} and references therein).
We will show that the Andreev-Bashkin drag is proportional to the multiparticle contribution to the first moment spin structure factor.
This results in a correction to the frequency of collective oscillations that could be measured in experiments to detect the drag.

\subsection{Superfluid densities as current-current response functions}\label{subsec:currentcurrent}

We take the approach described in \cite{Baym} and generalize it to a two component superfluid mixture.
The system considered here is described by an Hamiltonian in the form
$
\hat{H}=\hat{K}+\hat{U}
$
where the kinetic term is 
\begin{equation}
%\hat{K}=\int d\mathbf{r} \, \left(\frac{\mathbf{\hat{p}}_{A}^2(\mathbf{r})}{2m_A}+\frac{\mathbf{\hat{p}}_{B}^2(\mathbf{r})}{2m_B} \right)
\hat{K}=\sum_{\alpha=A,B}\int\left(\frac{\hbar^2}{2m_\alpha}\nabla\hat{\Psi}_\alpha^\dagger\nabla\hat{\Psi}_\alpha\right)d\mathbf{r}
\end{equation}
and the interaction reads
\begin{equation}
\hat{U}_{\alpha\beta}=\frac{1}{2}\!\sum_{\alpha,\beta}\int\! d \mathbf{r} d\mathbf{r}' \hat{\Psi}_\alpha^\dagger(\mathbf{r}) \hat{\Psi}_\beta^\dagger(\mathbf{r}') U_{\alpha\beta}(\mathbf{r}-\mathbf{r}') \hat{\Psi}_\beta(\mathbf{r}') \hat{\Psi}_\alpha(\mathbf{r}) \, , 
\end{equation}
with $U_{\alpha\beta}$ the two-body intra- and interspecies potential and $\hat{\Psi}_{\alpha}(\mathbf{r})$ and $\hat{\Psi}_{\alpha}^{\dagger}(\mathbf{r})$ quantum annihilation and creation fields of each species at position $\mathbf{r}$.
In order to define the superfluid densities microscopically, we will express the currents of Eq.~(\ref{AB}) as averages of the corresponding quantum operator:
\begin{equation}{
\hat{\mathbf{j}}_{\alpha}(\mathbf{r})  = \frac{\hbar }{2m_{\alpha}i}\left( \hat{\Psi}^{\dagger}_{\alpha} (\mathbf{r}) \nabla \hat{\Psi}_{\alpha}(\mathbf{r}) -H.c.  \right)},\label{currentop}\end{equation} namely we will ensure that $\left<\hat{\mathbf{j}}_{\alpha}(\mathbf{r}) \right> = \mathbf{j}_{\alpha}$.

To study the linear response of the system we consider the situation in which the superfluid is subject to a static transverse probe. The transverse probe can be thought of as an artificial gauge field described by a magnetic vector potential $\mathbf{A}(\mathbf{r})$ \cite{DalGer}.
We study the problem in the London gauge \cite{London} where the magnetic potential satisfies the condition:
\begin{equation} \label{eq:transversefield}
\mathbf{q} \cdot \mathbf{A}(\mathbf{q})=0 \, ,
\end{equation}
where $\mathbf{A}(\mathbf{q})$ is the Fourier transform of $\mathbf{A}(\mathbf{r})$.
In presence of this field the superfluid velocities $\mathbf{v}^{(s)}_{\alpha}$ read:
\begin{equation} 
\mathbf{v}_{\alpha}^{(s)}(\mathbf{r})=\frac{1}{2m_\alpha}\left(\nabla \varphi_\alpha -2e_\alpha \mathbf{A}(\mathbf{r})\right) \, .
\end{equation}
In the case of a charged superconductor $e_\alpha$ is the electrical charge, while for a neutral superfluid it can be simply thought of as a constant that quantifies the coupling of the artificial gauge field with the species $\alpha$.
In the ground state the condensate phase is a constant, thus $\nabla \varphi_\alpha=0$.

The hydrodynamic equations in Eq.~(\ref{AB}) for the current densities of the two species read:
\begin{equation}\label{AB2}
 m_\alpha \mathbf{j}_\alpha(\mathbf{r})= -\left(\rho_{\alpha \alpha} \frac{e_\alpha}{m_{\alpha}}+\rho_{\alpha \beta} \frac{e_\beta}{m_{\beta}} \right) \mathbf{A}(\mathbf{r}) \, .
\end{equation}
Eq.~(\ref{AB2}) is the London equation for a two component superfluid \cite{London}.

We will now compare this equation with the current response to a probe $\mathbf{A}(\mathbf{r})$ in order to identify the superfluid densities.
In presence of a transverse probe the current density operators transform in the following way:
\begin{equation}\label{eq:transformationj}
\hat{\mathbf{j}}_\alpha(\mathbf{r}) \rightarrow \hat{\mathbf{j}}_\alpha(\mathbf{r}) - \frac{e_\alpha}{m_\alpha} \hat{n}_\alpha(\mathbf{r}) \mathbf{A}(\mathbf{r})  \, ,
\end{equation}
where $\hat{n}_\alpha(\mathbf{r})=\hat{\Psi}_{\alpha}^{\dagger}(\mathbf{r})\hat{\Psi}_{\alpha}(\mathbf{r})$ is the density operator at position $\mathbf{r}$.
The Hamiltonian gets transformed to:
\begin{equation}\label{eq:transformationH}
\begin{split}
\hat{H}\rightarrow  \hat{H} & -\sum_{\alpha=A,B} \frac{e_\alpha}{m_\alpha} \int d \mathbf{r} \; \hat{\mathbf{j}}_{\alpha}(\mathbf{r}) \cdot  \mathbf{A}(\mathbf{r}) \\ & 
+\sum_{\alpha=A,B} \frac{e_\alpha^2}{2m_\alpha} \int d \mathbf{r} \; \hat{n}_{\alpha}(\mathbf{r})   A^2(\mathbf{r}) \, .
\end{split}
\end{equation}
Let's suppose without loss of generality that the vector potential $A$ is along the $x$ direction. 
We obtain the average of the current density operator to first order in the perturbation $\mathbf{A}(\mathbf{r})$:
\begin{equation}\label{Baymcurrent}
\begin{split}
   & \left<\hat{j}_{x,\alpha}(\mathbf{r})\right>=- n_\alpha(\mathbf{r}) \frac{e_\alpha}{m_\alpha} A(\mathbf{r}) \\& + \sum_{\beta=A,B} e_\beta  \int d \mathbf{r}'  \chi_{j_{x,\alpha},j_{x,\beta}}(\mathbf{r},\mathbf{r}')  A(\mathbf{r}) \, , 
\end{split}
\end{equation}
where $\chi_{j_{x,\alpha},j_{x,\beta}}(\mathbf{r},\mathbf{r}')$ are the static current-current response functions.

For a homogeneous system of volume $V$ and number of atoms $N$ we have that $n_\alpha(\mathbf{r})=n_\alpha=N/V$ and we can recast (\ref{Baymcurrent}) in terms of the Fourier transform of the current-current response function at $T=0$ \cite{StrPit}:
\begin{equation}\label{eq:defresponse}
\begin{split}
&\chi_{j_{x,\alpha},j_{x,\beta}}(\mathbf{q}) = \\& \frac{1}{V} 
\sum_{n\neq 0} \left(\frac{\braket{ n|j^{\dagger}_{x,\alpha}(\mathbf{q})|0 } \braket{ 0|j_{x,\beta}(\mathbf{q})|n }}{E_n-E_0
}\right. \\& \left. -\frac{\braket{ n|j_{x,\alpha}(\mathbf{q})|0 } \braket{0|j^{\dagger}_{x,\beta}(\mathbf{q})|n}}{E_0-E_n
} \right)\, ,
\end{split}
\end{equation}
where $\ket{n}$ and $E_n$ are respectively the eigenstates and eigenergies in absence of the velocity perturbation, and ${n=0}$ corresponds to the ground state.
The operator $\hat{\mathbf{j}}_{\alpha}(\mathbf{q})$ is the Fourier transform of the current density operator of Eq.~(\ref{currentop}):
\begin{equation}
\hat{\mathbf{j}}_{\alpha}(\mathbf{q})= \frac{\hbar}{2 m_\alpha } \sum_{\mathbf{k}}\left( 2\mathbf{k}+\mathbf{q} \right) \hat{a}^{\dagger}_{\mathbf{k},\alpha} \hat{a}_{\mathbf{k}+\mathbf{q},\alpha} \, , \label{eq:j(q)}
\end{equation}
with ${\hat{a}_{\mathbf{k},\alpha}=V^{-1/2}\int d \mathbf{r} \, e^{i\mathbf{k}\cdot \mathbf{r}}   \hat{\Psi}_{\alpha}(\mathbf{r})}$.
 We point out that the current response function defined in Eq.~(\ref{eq:defresponse}) is intensive.

Because of Eq.~(\ref{eq:transversefield}) the Fourier transform of Eq.~(\ref{Baymcurrent}) will give the \textit{transverse} response function \cite{PinNoz}.
For an arbitrary direction of $\hat{\mathbf{j}}_\alpha (\mathbf{q})$ we have the following definitions for the transverse and longitudinal response functions:
\begin{equation}\begin{split}
   & \chi_{\mathbf{j}_{\alpha},\mathbf{j}_{\beta}}(q^{T}, q^{L}=0)\equiv \chi_{\mathbf{j}_\alpha,\mathbf{j}_\beta}^{T}(\mathbf{q}) \, , \\
   &\chi_{\mathbf{j}_{\alpha},\mathbf{j}_{\beta}}(q^{T}=0, q^{L}  )\equiv \chi_{\mathbf{j}_\alpha,\mathbf{j}_\beta}^{L}(\mathbf{q}) \, , \label{limitcommute}
\end{split}
\end{equation}
where $q^{T}$ and $q^{L}$ are the components of $\mathbf{q}$ perpendicular and parallel to $\mathbf{j}$ respectively.
The linear response result for the current carried by each component reads:
\begin{equation}\label{eq:finallinearresponse}
  \left<\hat{\mathbf{j}}_\alpha \right>=- n_\alpha \frac{e_\alpha}{m_\alpha} \mathbf{A}  + \sum_{\beta=A,B}  e_\beta \chi^{T}_{j_{\alpha},j_{\beta}}(\mathbf{q}=0) \mathbf{A} \, ,
\end{equation}
with $\mathbf{A}\equiv \mathbf{A}(\mathbf{q}=0)$.

We can now match this result of linear response theory with the hydrodynamics predicted by Eq.~(\ref{AB2}).
Comparing Eq.~(\ref{AB2}) and Eq.~(\ref{eq:finallinearresponse}) we get the desired results for the superfluid densities in terms of linear response functions:
\begin{equation}\label{rhodrag1}
-m_A m_B \lim_{\mathbf{q} \to 0}\chi_{\mathbf{j}_A,\mathbf{j}_B}^{T}(\mathbf{q}) = \rho_{AB} \, ,
\end{equation}
\begin{equation}\label{rhodrag2}
m_{\alpha} n_{\alpha}-m_{\alpha}^2\lim_{\mathbf{q} \to 0}\chi_{\mathbf{j}_{\alpha},\mathbf{j}_{\alpha}}^{T}(\mathbf{q}) =   \rho_{\alpha\alpha} \, ,
\end{equation}
\begin{equation}\label{rhodrag3}
\sum_{\alpha,\beta=A,B} m_\alpha m_\beta \lim_{\mathbf{q} \to 0} \chi_{\mathbf{j}_{\alpha},\mathbf{j}_{\beta}}^{T}(\mathbf{q})=   \rho_n \, .
\end{equation}
Where the last equation can be obtained from the other two by using Eq.~(\ref{eq:normalization2}), and for a translational invariant system at zero temperature $\rho_n = 0$.

Note that in the single component superfluid, the zero temperature transverse response is zero, since the superfluid fraction is equal to the total mass density of the system. Here, we can also consider the relative response of $\hat{\mathbf{j}}_{B}(\mathbf{r})$  to $\hat{\mathbf{j}}_{A}(\mathbf{r}) \cdot  \mathbf{A}(\mathbf{r})$ i.e. $\chi^{T}_{\mathbf{j}_{A},\mathbf{j}_{B}}$ which is nonzero even in the groundstate as we see from (\ref{rhodrag1}). Nevertheless, the total response to a transverse field coupling to both currents,  $\chi^{T}_{\mathbf{j}_{A},\mathbf{j}_{A}}+\chi^{T}_{\mathbf{j}_{B},\mathbf{j}_{B}}+2\chi^{T}_{\mathbf{j}_{A},\mathbf{j}_{B}}$, remains zero.
In this sense the drag behaves as a sort of normal component at zero temperature decreasing the value of the diagonal superfluid densities $\rho_{\alpha \alpha}$ and taking part in the transverse response.

The formalisation of Eqs.~(\ref{rhodrag1}), (\ref{rhodrag2}), (\ref{rhodrag3}) is an important result of the paper. Resting only on the assumption of having a homogeneous superfluid mixture with interspecies interaction, they provide the full microscopic expressions of the hydrodynamic coefficients to linear order in the superfluid velocities.
In particular, Eq.~(\ref{rhodrag1}) describes the drag as a mutual correlation between currents in the two species.

In the case of a lattice described by a single-band Hubbard Hamiltonian Eqs.~(\ref{rhodrag1}), (\ref{rhodrag2}), (\ref{rhodrag3}) are slightly modified due to the lack of continuous translational invariance. As we show in Appendix \ref{subsec:latticedrag}, $\hat{n}_{\alpha}$ has simply to 
be replaced by the kinetic energy density $\hat{K}_{\alpha}$.

\subsection{Current response and sum rules}\label{sec:sumrules}
In a system where a single species is present the longitudinal current response function is proportional to the first moment of the structure factor (see, e.g., \cite{StrPit}) in the static limit.
In what follows we will extend this notion to the case of a two component system and relate it to the drag coefficient.

%First of all we will relate the f-sum rule for the structure factor to the current response functions%

The structure factor at zero temperature for an operator $\hat{F}$ is defined as:
\begin{equation}\label{eq:structurefactor}
S_F(\omega)= \sum_{n}  | \braket{n |\hat{F}| 0}|^2 \, \delta(\hbar \omega-\hbar \omega_{n0}) \, ,
\end{equation}
where $\hbar \omega_{n0}=E_n-E_0$.
We will be concerned here with the cases in which the operator $\hat{F}$ is the density operator (with corresponding structure factor $S_d$) 
\begin{equation}
\hat{\rho}_{\mathbf{q}}=\sum_{\mathbf{k}} \left(  \hat{a}^{\dagger}_{\mathbf{k}+\mathbf{q},A}\hat{a}_{\mathbf{k},A}+\hat{a}^{\dagger}_{\mathbf{k}+\mathbf{q},B}\hat{a}_{\mathbf{k},B}\right) \, ,
\end{equation} 
or the spin operator (respectively $S_s$) \begin{equation}\hat{s}_{\mathbf{q}}=\sum_{\mathbf{k}} \left(  \hat{a}^{\dagger}_{\mathbf{k}+\mathbf{q},A}\hat{a}_{\mathbf{k},A}-\hat{a}^{\dagger}_{\mathbf{k}+\mathbf{q},B}\hat{a}_{\mathbf{k},B}\right) \, .
\label{eq:spinop}
\end{equation}

Let us specialize here to the case where the two components have equal masses $m_A=m_B=m$ and densities $n_A=n_B=n$, the generalization being quite straightforward. The density structure factor satisfies the f-sum rule (see, e.g., \cite{PinNoz}, Vol. I, Chapter 4):
\begin{equation}\label{eq:m1d}
\begin{split}
M_{1,d}(\mathbf{q}) &=  \frac{1}{V}\int_0^\infty \omega S_{d}(\mathbf{q},\omega) \, d\omega  = \\ & =  \frac{1}{2V \hbar^2}\braket{\left[\hat{\rho}_{\mathbf{q}}^\dagger, \left[\hat{H},\hat{\rho}_{\mathbf{q}}\right] \right]}=\frac{n}{m} q^2 \, ,
\end{split}
\end{equation}
where we defined $M_{1,d}(\mathbf{q})$ the first moment of the density structure factor.
An analogous result is valid for the spin structure factor, namely:
\begin{equation}\label{eq:m1s}
\begin{split}
M_{1,s}(\mathbf{q}) & =  \frac{1}{V}\int_0^\infty \omega S_{s}(\mathbf{q},\omega) \, d\omega  = \\ & =  \frac{1}{2V \hbar^2}\braket{\left[\hat{s}_{\mathbf{q}}^\dagger, \left[\hat{H},\hat{s}_{\mathbf{q}}\right] \right]}=\frac{n}{m} q^2 \, ,
\end{split}
\end{equation}
where $M_{1,s}(\mathbf{q})$ is the first moment of the spin structure factor.

The first moment of the density (respectively spin) structure factor are related to longitudinal density (respectively spin) current response functions \cite{PinNoz}.
Consider in fact the density (respectively spin) current ${\mathbf{j}_{d(s)}(\mathbf{q})=\mathbf{j}_A (\mathbf{q}) \pm \mathbf{j}_B (\mathbf{q})}$ and its longitudinal response function:
\begin{equation}
\begin{split}
\chi^{L}_{\mathbf{j}_{d(s)}\mathbf{j}_{d(s)}}(\mathbf{q})& \equiv \chi_{\mathbf{j}_{A},\mathbf{j}_{A}}^{L}(\mathbf{q})+\chi_{\mathbf{j}_{B},\mathbf{j}_{B}}^{L}(\mathbf{q}) \pm 2\chi_{\mathbf{j}_{A},\mathbf{j}_{B}}^{L}(\mathbf{q}) \, . \label{eq:parallel}
\end{split}
\end{equation}
Using the definition of the structure factor, Eq.~(\ref{eq:structurefactor}), and the continuity equation we obtain:
\begin{equation}
\begin{split}
\label{eq:sumrule}
\chi^{L}_{\mathbf{j}_{d(s)}\mathbf{j}_{d(s)}}(\mathbf{q}) & =  \frac{2}{V q^2} \sum_n \frac{| \braket{0|\mathbf{q}\cdot \mathbf{j}_{d(s)}(\mathbf{q})|n} |^2}{E_n-E_0}=\\ & =\frac{2}{V q^2}\int_0^\infty \omega S_{d(s)}(\mathbf{q},\omega) d\omega= \frac{2}{q^2} M_{1,d(s)}(\mathbf{q}) \, .
\end{split}
\end{equation}
In order to find an expression for the drag in terms of the first moment of the structure factor we need now to relate the longitudinal response to the transverse response as it is the latter which is proportional to $\rho_{AB}$ by Eq.~(\ref{rhodrag1}).
To do so, it is necessary to separate the contributions of single particle and multiparticle excitations to the response functions, as they allow to discriminate between the longitudinal and transverse response.

At zero temperature the excited states that will contribute to the matrix elements of the response functions in Eqs.~(\ref{eq:m1d}), (\ref{eq:m1s}) and (\ref{eq:sumrule}) can be distinguished in two categories: single particle and multiparticle excitations \cite{PinNoz}.
The former are obtained from the ground state by adding a single quasiparticle of momentum $\mathbf{q}$ to the ground state, while the latter consist of several excited quasiparticles with total momentum $\mathbf{q}$.
In a single component superfluid, conservation of total current allows us to conclude that multiparticle excitations give a negligible contribution to the first moment of the structure factor at long wavelengths.

In two component superfluids the situation is different: multiparticle states excited by the spin operator are not negligible even at long wavelengths.
This is a consequence of the fact that, unlike the total current ${\mathbf{j}_{d}(\mathbf{q})}$, the spin current ${\mathbf{j}_{s}(\mathbf{q})}$ is not conserved \cite{Leggettspin}.

Indeed, conservation laws define the long wavelength behaviour of the matrix elements that appear in the f-sum rule (we refer here to the approach contained in \cite{PinNoz}).
For the density current we have that
\begin{equation}\label{eq:currentconservation}
\lim_{\mathbf{q} \rightarrow 0} \braket{0| \mathbf{j}_d (\mathbf{q}) | n} =0 \, ,
\end{equation}
when $\ket{n}$ is a multiparticle excited state, since the total current is a good quantum number.
Using the continuity equation we obtain: 
\begin{equation}\label{eq:continuity}
(\omega_n-\omega_0)\braket{0|\rho_{\mathbf{q}} (\mathbf{q}) | n}=-\braket{0| \mathbf{q} \cdot \mathbf{j}_d (\mathbf{q}) | n}\, .
\end{equation}
Since the frequency $\omega_n-\omega_0$ will be constant at long wavelengths for multiparticle excitations \cite{PinNoz} this allows to conclude that $\braket{0|\rho_\mathbf{q}|n}$ will tend to $0$ at least as fast as $q^2$.
A similar argument is not valid for the spin operator as the corresponding spin current is not conserved, so Eq.~(\ref{eq:currentconservation}) does not hold.

Further analysis of conservation laws allows us to completely identify the low $\mathbf{q}$ limit of the contributions to $M_{1,s}$ coming from $\braket{0|\hat{s}_{\mathbf{q}}|n}$ and $(\omega_{n}-\omega_0)$ for long wavelengths. These results are summarized in table \ref{table} with the corresponding results for the density operator.
\begin{table}[h!]
\centering
\begin{tabular}{|p{2cm}||c|c|}
 %\hline
 %\multicolumn{1}{|c|}{Sum Rule contributions} \\
 \hline
 & Single Particle &Multiparticle\\
 \hline
 $|\braket{0|\hat{\rho}_{\mathbf{q}}|n}|^2$ &$q$ &$q^4$ \\
 $| \braket{0|\hat{s}_{\mathbf{q}}|n}|^2$   & $q$    &$q^2$\\
 $(\omega_{n}-\omega_0)$&  $q$ & const.   \\
 $M_{1,d}$ & $q^2$ & $q^4$ \\
 $M_{1,s}$ & $q^2$ & $q^2$ \\
 
 \hline
\end{tabular} 
\caption{Contributions to the sum rule in the low $q$ limit.}
\label{table}
\end{table}

The table shows how, in the f-sum rule for the spin moment $M_{1,s}$, single particle and multiparticle excitations contribute at the same order in $\mathbf{q}$, while for the density moment $M_{1,d}$, single particle excitations are dominant and so they exhaust the sum rule in the long wavelength ($\mathbf{q} \to 0$) limit.

Since the longitudinal current response is connected to the first moment by Eq.~(\ref{eq:sumrule}), the result summarized in table \ref{table} implies that longitudinal current response is determined by single particle and multiparticle excitations in the spin channel, and by single particle excitations only in the density channel.
We will see in what follows that the situation is different for the transverse response function.

The correlation functions $\chi_{\mathbf{j}_\alpha,\mathbf{j}_\beta}^{T} (\mathbf{q})$ are determined in the low $\mathbf{q}$ limit by matrix elements which connect multiparticle excited states only. Single particle excitations cannot contribute to the transverse response because they have a defined axial symmetry determined by their momentum $\mathbf{q}$ (see \cite{PinNoz}, Vol. II, Chapter 4), and a transverse current cannot excite them.
On the other hand, multiparticle excitations do not possess any symmetry around the ${\bf q}$ axis  and can contribute to the transverse response $\chi_{\mathbf{j}_\alpha,\mathbf{j}_\beta}^{T} (\mathbf{q})$. 
Thus, we can write:
\begin{equation}
\lim_{\mathbf{q} \to 0}\chi_{\mathbf{j}_\alpha,\mathbf{j}_\beta}^{T} (\mathbf{q})=\lim_{\mathbf{q} \to 0}\chi_{\mathbf{j}_\alpha,\mathbf{j}_\beta}^{T} (\mathbf{q})^{(m.p.)} \, .
\end{equation}
With the superscript ``m.p." we indicate that the only contributions are matrix elements connecting the ground state to multiparticle excited states.

A nonzero response of the superfluid to a transverse probe is thus strictly connected with the presence of multiparticle excitations in the multicomponent system, while in the case of a single component single particle excitations are the only low lying excited states at zero temperature and cannot be excited by such a probe.
This important result will also be used in calculating the drag for a weakly interacting Bose mixture in subsec.~\ref{subsec:bogoliubov}. There, only matrix elements which connect excited states made of a density and a spin phonon will appear in the transverse response.

The same reasoning does not apply to the longitudinal response where single particle excitations have the same axial symmetry as the probe. Nevertheless we should expect that the contribution of multiparticle excitations to the transverse and the longitudinal response is the same as they do not have any axial symmetry that can discriminate between the two. We can write that:
\begin{equation}\label{tran=long}
\lim_{\mathbf{q} \to 0}\chi_{\mathbf{j}_\alpha,\mathbf{j}_\beta}^{T} (\mathbf{q}) ^{(m.p.)}=\lim_{\mathbf{q} \to 0}\chi_{\mathbf{j}_\alpha,\mathbf{j}_\beta}^{L} (\mathbf{q})^{(m.p.)}\, .
\end{equation}
Using Eqs.~(\ref{tran=long}) and (\ref{rhodrag1}) we can now write $\rho_{AB}$ in terms of the longitudinal response:
\begin{equation} \label{eq:dragdensityspin}
\frac{\rho_{AB}}{m^2}
=-\lim_{\mathbf{q} \to 0}\chi_{\mathbf{j}_\alpha,\mathbf{j}_\beta}^{T} (\mathbf{q}) ^{(m.p.)}=-\lim_{\mathbf{q} \to 0} \chi_{\mathbf{j}_{A},\mathbf{j}_{B}}^{L}(\mathbf{q})^{(m.p.)} .
\end{equation}
This equation allows the drag to be expressed directly from the multiparticle contributions to the sum rule
$M_{1,s}$. Using Eqs.~(\ref{eq:parallel}), (\ref{eq:sumrule}) and (\ref{eq:dragdensityspin}) one has:
\begin{equation} \label{eq:multimagnon}
\begin{split}
\frac{\rho_{A B}}{m^2} & =4\left(\chi^{L}_{\mathbf{j}_{d}\mathbf{j}_{d}}(\mathbf{q})^{(m.p.)} - \chi^{L}_{\mathbf{j}_{s}\mathbf{j}_{s}}(\mathbf{q})^{(m.p.)}\right)\\&
=\lim_{\mathbf{q} \to 0}\frac{1}{2q^2}M^{(m.p.)}_{1,s }(\mathbf{q}) \, ,
\end{split}
\end{equation}
where we used the fact that the density sum rule is exhausted by single particle excitations in the low $\mathbf{q}$ limit, namely:
\begin{equation}
    \lim_{\mathbf{q} \to 0}\frac{2}{q^2}M^{(m.p.)}_{1,d}(\mathbf{q})=0 \, .
\end{equation}
Finally making use of the last equality in Eq.~(\ref{eq:m1s}):
\begin{equation}\label{eq:dragsumrule}
M_{1,s}^{(s.p.)}=\frac{nq^2}{m}\left(1-\frac{2\rho_{AB}}{mn} \right) \, .
\end{equation}

The superscript $(s.p.)$ indicates the fact that we are accounting for contributions to $M_{1,s}$ coming only from single particle excitations.
Eq.~(\ref{eq:dragsumrule}) expresses a crucial result for a two species superfluid: in the low $\mathbf{q}$ limit the single particle contribution to $M_{1,s}$ is reduced from the value predicted by the f-sum rule by a factor proportional to the drag ${\rho_{AB}}$, which accounts for multiparticle excitations.
In a single component superfluid this contribution is absent and the response of the system in the static, long wavelength limit is accounted for by single particle excitations only.

By computing the single particle contribution to the first moment of the spin operator, using Eq.~(\ref{eq:dragsumrule}), we are able to estimate the frequency of collective spin excitations, as we will do in subsec.~\ref{sec:sound}, and show how the drag influences them.

The presence of a finite drag $\rho_{AB}$ implies that the low energy, long wavelength quantum hydrodynamic Hamiltonian for two superfluids contains off-diagonal superfluid densities (see, e.g., \cite{NesRec}): 
\begin{equation}\label{hydrohamiltonian}
H_{eff}=\!\!\!\!\sum_{\alpha, \beta= A,B}\int d\mathbf{r} \left(  \frac{\hbar^2\rho_{\alpha \beta}}{2m^2} \nabla \hat{\phi}_{\alpha} \cdot \nabla \hat{\phi}_{\beta}+\frac{g_{\alpha\beta}}{2}\hat{\Pi}_\alpha \hat{\Pi}_\beta\right) \, .
\end{equation}
In the previous expression $\hbar \nabla\hat{\phi}_\alpha/m_\alpha$ is the superfluid velocity fluctuation of component $\alpha$, $\hat{\Pi}_\alpha$ the density fluctuation and $g_{\alpha\beta}=\partial^2\epsilon/\partial n_\alpha\partial n_\beta$ with $\epsilon$ the ground state energy is the compressibility matrix.
The operators $\hat{\phi}_{\alpha}$ and $\hat{\Pi}_{\alpha}$ satisfy the commutation relations: ${\left[\hat{\phi}_{\alpha}(\mathbf{r}),\hat{\Pi}_{\beta}(\mathbf{r}') \right]=i\hbar \delta_{\alpha \beta} \delta (\mathbf{r}-\mathbf{r}')}$.
Using the effective Hamiltonian 
%the single particle contribution to the first moment $M_{1,F}^{(s.p.)}$
the first moment of an operator $\hat{F}$ can be also calculated \cite{StrPit} {\textsl via} the commutator:
\begin{equation}\label{eq:m1eff}
M_{1,F}^{eff} =\frac{1}{2V \hbar^2}\braket{\left[\hat{F}^\dagger, \left[\hat{H}_{eff},\hat{F}\right] \right]}. 
\end{equation}
In particular for a spin density perturbation where ${\hat{F}=\hat{\Pi}_{A,\mathbf{q}}-\hat{\Pi}_{B,\mathbf{q}}}$ and indeed Eq.~(\ref{eq:m1eff}) coincides with the single particle excitation result Eq.~(\ref{eq:dragsumrule}).

%Since single particle excitations exhaust the density sum rule, the first moment of the density structure factor satisfies:
%\begin{equation}\label{eq:densitymoment}
%M^{(s.p.)}_{1,d }(\mathbf{q})=M_{1,d }(\mathbf{q})=\frac{m_A n_A+m_B n_B}{2m_A m_B} q^2 \, .
%\end{equation}

\section{Weakly interacting Bose-Bose mixture: drag and excitations}\label{sec:WIBG}
In this section we put the linear response formalism to test in the case of a weakly interacting Bose-Bose mixture.
In subsec.~\ref{subsec:bogoliubov} as a benchmark for our formalism we compute the drag in the weakly interacting mixture reproducing a known result \cite{FilShe}.
Then in subsecs.~\ref{sec:sound} and \ref{sec:dipole} we make use of sum rules to compute the effect of the drag on collective excitations. 
We calculate the full beyond mean field correction to the frequencies of spin waves and of the spin dipole collective mode.
The drag results in a change in the frequency of excitations in the spin channel.

Even though we focus on the specific case of a weakly interacting Bose-Bose mixture, the procedure we take in this section can be applied in principle to other systems where the effect could be more sizeable.
\subsection{Collisionless drag for a weakly interacting Bose-Bose mixture}\label{subsec:bogoliubov}
A strength of the linear response formalism we developed in subsec.~\ref{subsec:currentcurrent} is that it provides a method to compute easily the superfluid density matrix.
To show it at work we now compute the relevant response functions in the Bogoliubov approximation for a weakly interacting Bose-Bose mixture at zero temperature.
With the computations in this subsection we reproduce in a very easy way the results in Ref. \cite{FilShe}. For completeness in Appendix \ref{subsec:latticedrag} we also show how to easily recover the result for the Hubbard model obtained in \cite{LinSub}. 

We will compute the current response in one component to a probe current in the other component along the $x$ direction. Since the system is isotropic, the response will also be along the $x$ direction.
For simplicity we will assume a $\mathbb{Z}_2$ symmetric mixture, meaning that the densities, the masses and the intraspecies contact interactions of the two components are equal, namely $n_A=n_B=n$, $m_A=m_B=m$ and $g_{AA}=g_{BB}=g$.
Additionally, at zero temperature, Eq.~(\ref{rhodrag3}) for the $\mathbb{Z}_2$ symmetry implies the following useful equality:
\begin{equation}
\chi^{T}_{j_{x,A},j_{x,A}}(\mathbf{q}, \omega)=\chi^{T}_{j_{x,B},j_{x,B}}(\mathbf{q}, \omega)=-\chi^{T}_{j_{x,A},j_{x,B}}(\mathbf{q}, \omega). \label{eq:helpfulequality}
\end{equation}

%\subsubsection{Drag in a uniform mixture}
We consider the case of a homogeneous weakly interacting Bose-Bose mixture with volume $V$.
The mixture is stable in the mean field approximation when $g>0$ and $|g_{AB}|<g$, where $g_{AB}$ is the interspecies coupling. 
Since linear response requires the ground state to be stable we will only consider the case $|g_{AB}|<g$.
The Hamiltonian describing the system, written in the momentum space basis, is:
\begin{equation}
\begin{split}\label{eq:hamiltonian}
\hat{H} & =\sum_{\alpha=A,B}  \sum_\mathbf{k}  \epsilon_{\mathbf{k}} \hat{a}^{ \dagger}_{\mathbf{k},\alpha} \hat{a}_{\mathbf{k},\alpha} \\ &
+\frac{g}{2V} \sum_{\alpha=A,B} \sum_{\mathbf{k}_1, \mathbf{k}_2, \mathbf{p}}  \hat{a}^{ \dagger}_{ \mathbf{k}_1+, \alpha}  \hat{a}^{ \dagger}_{ \mathbf{k}_2-\mathbf{p}, \alpha} \hat{a}_{ \mathbf{k}_1,\alpha} \hat{a}_{ \mathbf{k}_2,\alpha} \\&  +
\frac{g_{AB}}{2V} \sum_{\mathbf{k}_1, \mathbf{k}_2, \mathbf{p}}  \hat{a}^{ \dagger}_{ \mathbf{k}_1+\mathbf{p}, A}  \hat{a}^{ \dagger}_{\mathbf{k}_2-\mathbf{p}, B} \hat{a}_{\mathbf{ k}_1,A} \hat{a}_{ \mathbf{k}_2,B} \, ,
\end{split}
\end{equation}
where $\hat{a}_{ \mathbf{k}, \alpha}$ and
$\hat{a}^{ \dagger}_{ \mathbf{k}, \alpha}$ respectively annihilate and create a particle of species $\alpha$ and momentum $\mathbf{k}$ and ${\epsilon_{\mathbf{k}}=\hbar^2 k^2/2m}$.

The intraspecies and interspecies couplings are related to the scattering lengths by $g=4\pi \hbar^2 a/m$ and ${g_{AB}=4\pi \hbar^2 a_{AB}/m}$ respectively.
In the weakly interacting limit, when $na^3 \ll 1$, quantum fluctuations are small and the Hamiltonian in Eq.~(\ref{eq:hamiltonian}) can be reduced to a quadratic form by means of the Bogoliubov approximation. 
We retain only terms which are quadratic in the operators $a_{\mathbf{p},\alpha}$ and $a_{\mathbf{p},\alpha}^\dagger$ for $\mathbf{p}\neq 0$  and replace $a_{\mathbf{p}=0,\alpha}$ and $a^{\dagger}_{\mathbf{p}=0,\alpha}$ with $\sqrt{N}_0$, where $N_0$ is the number of particles in the condensate.
The quadratic Hamiltonian can be diagonalized by a canonical transformation to the basis of Bogoliubov quasiparticles $\hat{b}^{\dagger}_{d,\mathbf{k}}$ and $\hat{b}^{\dagger}_{s,\mathbf{k}}$ in a balanced two component mixture \cite{TomDep}:
\begin{equation}\label{Bogo}\begin{split}
&
\hat{a}_{\mathbf{k},A}=\frac{1}{\sqrt{2}}(u_{d,\mathbf{k}} \hat{b}_{d,\mathbf{k}} + v_{d,\mathbf{k}}\hat{b}^{\dagger}_{d,-\mathbf{k}} + u_{s,\mathbf{k}} \hat{b}_{s,\mathbf{k}} + v_{s,\mathbf{k}} \hat{b}^{\dagger}_{s,-\mathbf{k}}) \, ,
\\&
\hat{a}_{\mathbf{k},B}=\frac{1}{\sqrt{2}}(u_{d,\mathbf{k}} \hat{b}_{d,\mathbf{k}} + v_{d,\mathbf{k}}\hat{b}^{\dagger}_{d,-\mathbf{k}} - u_{s,\mathbf{k}} \hat{b}_{s,\mathbf{k}} - v_{s,\mathbf{k}} \hat{b}^{\dagger}_{s,-\mathbf{k}}) \, .
\end{split}
\end{equation}
The labels $d$ and $s$ indicate density and spin quasiparticles and the coefficients $u_{d(s),\mathbf{k}}$ and $v_{d(s),\mathbf{k}}$ are:
\begin{equation}
\begin{split}
& u_{d(s),\mathbf{k}}=\frac{1}{2} \left(\sqrt{\frac{\epsilon_\mathbf{k}}{\Omega_{d(s),\mathbf{k}}}}+\sqrt{\frac{\Omega_{d(s),\mathbf{k}}}{\epsilon_\mathbf{k}}} \right) \, ,
\\ &
v_{d(s),\mathbf{k}}=\frac{1}{2} \left(\sqrt{\frac{\epsilon_\mathbf{k}}{\Omega_{d(s),\mathbf{k}}}}-\sqrt{\frac{\Omega_{d(s),\mathbf{k}}}{\epsilon_\mathbf{k}}} \right) \, ,
\end{split}
\end{equation}
where $\Omega_{d,\mathbf{k}}$ and $\Omega_{s,\mathbf{k}}$ are the excitation energies of the density and spin excitations respectively, namely:
\begin{equation}
\Omega_{d (s),\mathbf{k}}=\sqrt{\frac{\hbar^2 k^2}{2m} \left (\frac{\hbar^2 k^2}{2m}+2 g n\pm 2g_{\text{AB}} n \right)}.
\label{eq:spectrum}
\end{equation}
The diagonalized Hamiltonian takes the form:
\begin{equation}
\hat{H} \approx E_0+\sum_{\gamma=d,s} \sum_{\mathbf{k} \neq 0}\Omega_{\gamma,\mathbf{k}}\hat{b}^{\dagger}_{\gamma,\mathbf{k}}\hat{b}_{\gamma,\mathbf{k}} \,  ,
\end{equation}
where $E_0$ is the ground state energy.
We substitute the above expressions into Eqs.~(\ref{rhodrag1}), (\ref{rhodrag2}) and (\ref{rhodrag3}) and use Eqs.~(\ref{eq:j(q)}) and (\ref{eq:helpfulequality}) to find the expression of the drag in terms of the excitation spectra $\Omega_{d (s),\mathbf{k}}$:
\begin{equation}\label{final}
\begin{split}
\rho_{
AB} &= -m^2\lim_{\mathbf{q} \to 0}\chi^{T}_{j_{x,A},j_{x,B}}(\mathbf{q}) \\& =\frac{\hbar^2}{2V}\sum_{\mathbf{k}} \frac{(u_{d,\mathbf{k}}v_{s,\mathbf{k}}-u_{s,\mathbf{k}}v_{d,\mathbf{k}})^2}
{\Omega_{d,\mathbf{k}}+\Omega_{s,\mathbf{k}}}k_x^2
\\& =\frac{\hbar^2}{8V} \sum_{\mathbf{k}} 
 \frac{(\Omega_{d,\mathbf{k}}-\Omega_{s,\mathbf{k}})^2}{(\Omega_{d,\mathbf{k}}+\Omega_{s,\mathbf{k}})\Omega_{s,\mathbf{k}}\Omega_{d,\mathbf{k}}} k_x^2 \, .
 \end{split}
\end{equation}

This coincides with the result obtained in \cite{FilShe} where also the finite temperature result was derived.
The expression of Eq.~(\ref{final}) is rather suggestive.
The second line of the equation displays a result that we anticipated in subsec.~\ref{sec:sumrules} and that holds in general: only matrix elements between multiparticle excited states ($\sim b^\dagger b^\dagger |0 \rangle$) contribute to the transverse response, as expected from the discussion in subsec.~\ref{sec:sumrules}.
In the present case of a $\mathbb{Z}_2$ symmetric mixture the multiparticle excitations are composed of spin and density phonons.

The numerator on the third line of Eq.~(\ref{final}) implies that the collisionless drag strictly depends on the difference in the bare excitation energies in the spin and density channel.
The result of Eq.~(\ref{final}) is an even function of the interspecies interaction $g_{AB}$, as a result of the Bogoliubov approximation, while the inclusion of higher order terms should eliminate this symmetry.
In particular one can expect the drag to be stronger in the attractive regime where density-density fluctuations are enhanced.

The sum in Eq.~(\ref{final}) can be turned into an integral which can be solved analytically, giving:
\begin{equation}\label{eq:dragF}
\rho_{AB}=mn \sqrt{na^3} \eta^2 F(\eta) \, ,
\end{equation}
where $\eta=\frac{|g_{AB}|}{g}$ and:
\begin{equation}\label{eq:Ffunction}
F(\eta)=\frac{256}{45\sqrt{2\pi}}\cfrac{2+3\sqrt{(1+\eta) (1-\eta)}}{(\sqrt{2(1-\eta)}+\sqrt{2(1+\eta)})^3}  \, .
\end{equation}
%\begin{figure}[h!]
%\includegraphics[width=0.5\textwidth]{drag.pdf}
%\caption{Superfluid drag as a function of the interspecies interaction strength $\eta=|g_{AB}|/g$. The drag is seen to increase monotonically with $\eta$ to a maximum of $mn\sqrt{na^3}64/45\sqrt{2 \pi}$. Note also that $\rho_{AB}$ is an even function of $g_{AB}$ and so it has the same behaviour for attractive and repulsive interactions.}
%\end{figure}
We see from Eq.~(\ref{eq:dragF}) that the drag coefficient is directly proportional to the gas parameter $\sqrt{na^3}$, making the effect very small for weakly interacting mixtures.

A couple of remarks are worth.
In presence of a superfluid phase we would expect the response functions $\chi_{j_{x,A},j_{x,A}}(\mathbf{q})$ and $\chi_{j_{x,B},j_{x,B}}(\mathbf{q})$ to converge to the transverse (longitudinal) response when $q_x$ goes to zero before (after) $q_y$ and $q_z$ (Eq.~(\ref{limitcommute})).
However within the Bogoliubov approximation all the long wavelength limits commute, implying incorrectly that there is no distinction between the transverse and longitudinal response.
While the prediction for the transverse response is valid, the Bogoliubov approximation gives an incorrect value for the longitudinal response functions.
This is a known shortcoming of the Bogoliubov approximation, present also in the single component case, that can be cured by taking into account vertex corrections (see \cite{Schrie} for a detailed treatment).

%Also a different ordering of the $\omega=0$ and $\mathbf{q}=0$ limits produces in general different results.
%While the static response (obtained setting $\omega=0$ before ${\mathbf{q}=0}$) is related to the superfluid densities, the dynamic response in the low frequency limit (obtained setting ${\omega=0}$ after $\mathbf{q}=0$) allows us to calculate the Drude weight \cite{ScaWhi}, characterizing a zero viscosity ground state even in the absence of superfluidity.
%A finite Drude weight implies that the ground state of the system has zero viscosity or resistance, even in absence of superfluidity.
%At zero temperature the Drude weight and the superfluid weight coincide whenever the latter is finite \cite{ScaWhi}. This result is predicted correctly by the Bogoliubov theory, as the $\mathbf{q}=0$ and $\omega=0$ limits of the response functions commute when computed within this approximation.
%It would be interesting to consider the different case of a non-superfluid, two component perfect conductor and check wether the limit for $\omega=0$ and $\mathbf{q}=0$ commute in $\chi_{j_{x,1},j_{x,2}}(\mathbf{q}, \omega)$. 

\subsection{Beyond mean field correction to the spin speed of sound in a homogeneous gas}\label{sec:sound}
A crucial task in order to make the Andreev-Bashkin effect measurable is to predict its effect on physical observables that are accessible to experiments.
Here we address this point by computing the beyond mean field correction to the spin speed of sound. 
Importantly for the ongoing experiments on spin superfluidity (see, e.g., \cite{Ferrari2018,KimShi}) we show that for a weakly interacting repulsive Bose-Bose mixture the beyond mean field corrections to the spin speed of sound are dominated by the change in the susceptibility, while the collisionless drag gives a minor contribution. 
In the next subsec.~(\ref{sec:dipole}) we show instead that the two contributions are of the same order for the spin dipole mode frequency for a trapped mixtures.

For a homogeneous $\mathbb{Z}_2$ symmetric superfluid mixture the sounds that can propagate are the density and spin sound. While the former is not affected by the drag, the latter receives a correction at first order in the gas parameter $\sqrt{na^3}$.

In order to estimate the excitation energies we use sum rules for which we provided a number of results in the previous sections. 
In general indeed, within linear response, the energy of the lowest state excited by an operator $\hat{F}$ satisfies the inequality \cite{StrPit} 
\begin{equation}
\hbar \omega_s \leq \sqrt{\frac{M_{1,s}}{M_{-1,s}}} \, ,
\end{equation}
with $M_{\pm 1,s}$ the energy weighted and inverse energy weighted $M_{-1,s}=\int_0^\infty S_{s}(\mathbf{q},\omega)/\omega d\omega$ sum rules for the structure factor of the operator $\hat{F}$.  

\begin{figure}[h!]
\includegraphics[width=0.5\textwidth]{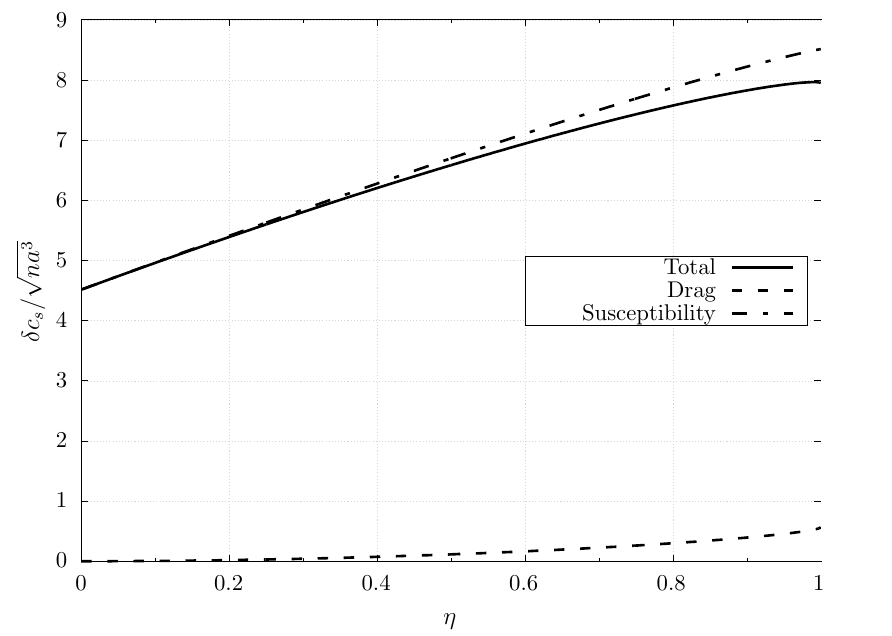}
\caption{Beyond mean field correction to the spin speed of sound (full line) as a function of the interspecies interactions $\eta=|g_{AB}|/g$.
We  limited the analysis to values of $\eta$ smaller than $1$ to avoid the regimes of phase separation and collapse.
The correction coming from the drag (dashed) is at least one order of magnitude smaller than that coming from Lee-Huang-Yang terms in the susceptibility (dotted).}\label{fig:speedofsound}
\end{figure}

By choosing as exciting operator the spin density operator Eq.~(\ref{eq:spinop}) we can estimate the spin dispersion relation. 
We use for the the energy weighted sum rules the general expression Eq.~(\ref{eq:dragsumrule}) with the drag given by the weakly interacting result Eq.~(\ref{eq:dragF}):
\begin{equation}
M_{1,s}^{(s.p.)}=\frac{nq^2}{m}\left( 1-\sqrt{na^3} \, 2\eta^2 F(\eta)  \right) \, .
\end{equation}
The inverse energy weighted moment, $M_{-1,s}$, is exhausted by single particle excitations as evident from table \ref{table} and it is given by:
\begin{equation}
M_{-1,s}=M_{-1,s}^{(s.p.)}=\frac{\chi_s}{2} \, ,
\end{equation}
where $\chi_s$ is the spin susceptibility, defined as ${\chi_s^{-1}= \frac{\partial^2(E/V)}{\partial(n_A-n_B)^2}}$, and $E$ is the energy of the superfluid in its ground state.
The latter has to be determined to the same order of $M_{1,s}$ and therefore one must use the equation of state for the Bose-Bose mixtures including the Lee-Huang-Yang correction \cite{Larsen1963}. Eventually we find  (see Appendix \ref{app:thermodynamics} for the derivation): 
\begin{equation}
\frac{1}{M_{-1,s}}= \left(g-g_{AB} \right) \left(1+ \sqrt{na^3} C(\eta) \right) \, ,
\end{equation}
where the function $C(\eta)$ is defined in Eq.~(\ref{eq:C}).
So we have that
\begin{equation}
 %\lim_{\mathbf{q}\to 0}
 \frac{M_{1,s}^{(s.p.)}}{M_{-1,s}}=c_s^2 q^2=c_{s,MF}^2[1+\sqrt{na^3}(C(\eta)-2\eta^2F(\eta))] q^2\, ,
\end{equation}
where $c_{s,MF}=\sqrt{n(g-g_{AB})/m}$ is the mean field spin speed of sound. 
In Fig. (\ref{fig:speedofsound}) we report the quantity:
 \begin{equation}
 \delta c_s=\cfrac{ c_s-c_{s,MF}}{c_{s,MF}}
 \end{equation}
which quantifies the deviation of the spin speed of sound from its mean field value. 

From Fig. (\ref{fig:speedofsound}) it is clear that the beyond mean field correction to the spin speed of sound is dominated by the susceptibility contribution. The overall correction for typical values of the gas parameter $\sqrt{na^3} \approx 10^{-3}$ is of order $10^{-3} - 10^{-2}$ but the contribution from the drag is at least one order of magnitude smaller.
We mention that a recent experiment \cite{KimShi} has been able to measure the two (density and spin) sounds of a symmetric superfluid mixture, although with the current precision it is not possible to observe beyond mean field effects on the spin speed of sound.

\subsection{Spin dipole modes in a trap}\label{sec:dipole}
We can apply a similar reasoning to the determination of the frequency of dipole modes for a Bose-Bose mixture trapped by a spherical harmonic potential:
\begin{equation}
V(r)=\frac{m}{2}\omega_0^2r^2 \, .
\end{equation}
assuming that it is in the Thomas-Fermi limit.
For a single component condensate the dipole mode corresponds to the oscillations of the center of mass of the system, characterized by the frequency $\omega_{D}= \omega_0 $
\cite{String}, independently of the interactions.
In the case of a two component mixture the dipole mode splits into two modes corresponding to the in and out of phase oscillations of the clouds of each component, corresponding to the density and spin modes respectively.
While for the in phase oscillations one obviously recovers the same result as for the single component Bose gas, the out of phase oscillations (i.e. the spin dipole mode) should be modified by beyond mean field corrections, as in the analogous case of Fermi-Liquid theory \cite{RecStr}.
 
We apply the formalism of sum rules to the spin dipole operator $\hat{F}=\hat{D}_s$:
\begin{equation}
\hat{D}_s=\int d^3\mathbf{r} \, x \, (\hat{n}_A(\mathbf{r})- \hat{n}_B(\mathbf{r}))\, .
\end{equation}
Again the spin dipole frequency can be estimated as:
\begin{equation}
\hbar \omega_{D_s}\leq \sqrt{\frac{M_{1,D_s}}{M_{-1,D_s}}} \, .
\end{equation}

Similarly to the case of the spin operator in the previous subsection, the single particle contribution to $M_{1,D_s}^{(s.p.)}$ is modified by the presence of the collisionless drag  as is evident from evaluating the commutator (Eq.~(\ref{eq:m1eff})):
\begin{equation}\label{eq:m1trap}
\begin{split}
M_{1,D_s}^{(s.p.)} & =\frac{1}{2V}\braket{\left[\hat{D}_s, \left[\hat{H}_{eff},\hat{D}_s\right] \right]} \\ &=  \frac{1}{V}\int d^3\mathbf{r}\frac{n(\mathbf{r})}{m}\left( 1-\frac{2\rho_{A B}(\mathbf{r})}{mn(\mathbf{r})}\right) \, .
\end{split}
\end{equation}
where the Hamiltonian $\hat{H}_{eff}$ is the effective hydrodynamic Hamiltonian of Eq.~(\ref{hydrohamiltonian}) with the addition of the harmonic trapping potential. The local drag coefficient $\rho_{A B} (\mathbf{r})$ is a function of the coordinates $\mathbf{r}$ only through $n(\mathbf{r})$, i.e. the local density approximation of Eq.~(\ref{eq:dragF}).
The inverse energy weighted moment, $M_{-1,D_s}$, is determined by the spin susceptibility via the expression \cite{Sartori}:
\begin{equation}\label{eq:m-1trap}
M_{-1,D_s}=\frac{1}{2V}\int d^3 \mathbf{r}\, x^2\chi_s(n(\mathbf{r})) \, .
\end{equation}
Consistently the density profile is determined within the local density approximation of the Lee-Huang-Yang equation of state in presence of the harmonic potential (see Appendix \ref{app:thermodynamics} for the derivation)\footnote{The use of the local density approximation for  $\rho_{AB}$ and $\chi_s$ is valid as long as the spin healing length $\xi_s =\sqrt{8\pi n \left(a-a_{AB}\right)}$ is much smaller than the size of the Bose-Bose mixture cloud, which we can identify with the Thomas-Fermi radius $R_{TF}$ (see Eq.~(\ref{eq:thomasfermi})).}.

%Again at mean field level the drag is absent and the susceptibility is just $\chi_{s,MF}= \frac{2}{g-g_{AB}}$, giving the following result for the mean field spin dipole frequency $\omega_{MF}$:
%\begin{equation}
%\omega_{MF}^2=\frac{g-g_{AB}}{g+g_{AB}} \omega_0^2
%\end{equation}
Eventually we  are able to write the sum rules as:
\begin{equation}\label{eq:m1m-1trap}
\begin{split}&
M_{1,D_s}^{(s.p.)}=\frac{n}{m}\left(1-\frac{15 \pi}{32} \sqrt{n(0)a^3} \eta^2 F(\eta) \right)  \, ,
\\
&
\frac{1}{M_{-1,D_s}}= \frac{5(g-g_{AB})}{R_{TF}^2}\left(1+\frac{5\pi}{32}\sqrt{n(0)a^3}(C(\eta)-B(\eta))\right) \, ,
\end{split}
\end{equation}
where $R_{TF}$ is the so-called mean-field Thomas-Fermi radius \cite{StrPit} of the trapped mixture:
\begin{equation} \label{eq:thomasfermi}
R_{TF}=\left(\frac{15N \left(g+g_{AB} \right)}{8 \pi m \omega_0^2}\right)^{1/5} \, .
\end{equation}
The expression for the functions $C(\eta)$ and $B(\eta)$ can be found in Appendix \ref{app:thermodynamics}, Eqs.~(\ref{eq:C}) and (\ref{eq:B}) respectively.

\begin{figure}[h!]
\includegraphics[width=0.5\textwidth]{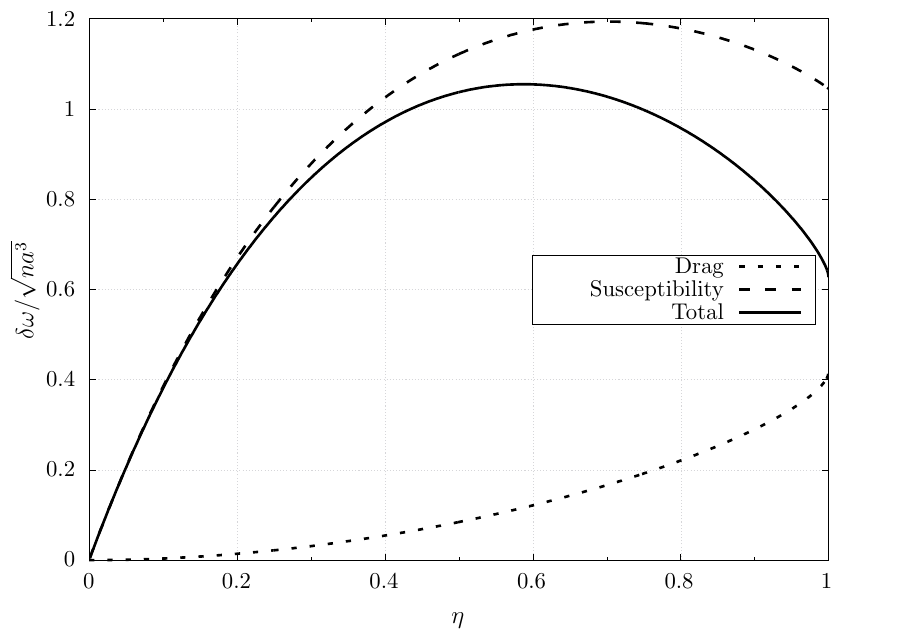}
\caption{
Beyond mean field correction to the spin dipole frequency (full line) as a function of the interspecies interactions $\eta=|g_{AB}|/g$.
We limited the analysis to values of $\eta$ smaller than $1$ to avoid the regimes of phase separation and collapse.
Although the overall correction is smaller than in the case of the spin speed of sound, here the effect of the drag (dashed) is comparable with that of the susceptibility (dotted).}\label{fig:spindipole}
\end{figure}

As for the homogeneous case we report in  Fig.(\ref{fig:spindipole}) the variation 
\begin{equation}
{\delta \omega=\cfrac{ \omega_{D_S}-\omega_{MF}}{\omega_{MF}}} \, ,
\end{equation}
which quantifies the deviation of the spin dipole frequency from its mean field value ${\omega_{MF}=\omega_0 \sqrt{(g-g_{AB})/(g+g_{AB})}}
$ and is always positive.
Also in the trapped cases the correction to the spin dipole frequency due to the drag is extremely small: for typical values of the gas parameter $\sqrt{n(0)a^3} \approx 10^{-3}$ the relative correction given by the drag is of order $10^{-3}- 10^{-4}$.
However, differently from the case of the speed of sound the correction from the drag can become comparable to the susceptibility one, owing to the non monotonic behaviour of the latter.
The correction coming from the susceptibility is not a monotonic function of $\eta$ (contrarily to the homogeneous case) as a result of the competition between the functions $C(\eta)$ and $B(\eta)$. The former comes from the beyond mean field correction to the spin susceptibility (Eq.~(\ref{eq:susceptibility})), the latter comes from the beyond mean field correction to the size of the cloud (Eq.~(\ref{eq:thomasfermi2})).
Finally notice that for $\eta=0$ the two clouds oscillate independently and we get indeed $\omega_{D_s}=\omega_0$ as for the in phase oscillations.

The calculations of this section can be easily extended to an elongated trap. For an ellipsoidal trap with frequencies $\omega_x$, $\omega_y$ and $\omega_z$ along the $x$, $y$ and $z$ direction Eq.~(\ref{eq:m1trap}) is unchanged while  Eq.~(\ref{eq:m-1trap}) has to be multiplied by a factor $\omega_x/\bar{\omega}_0$ where $\bar{\omega}_0=\sqrt[3]{\omega_x \omega_y \omega_z}$. The relative size of the correction from the susceptibility and the drag is unchanged.

The major drawback of weakly interacting gases is apparent from our results. The Andreev-Bashkin effect is made elusive by the fact that, being due to beyond mean field quantum fluctuations, it is typically very small in cold gases settings.
While increasing the strength of interactions would increase the magnitude of the drag, it would also amplify three-body losses.
We also expect that attractive interspecies interactions ($g_{AB}<0$) would enhance the effect as they increases density fluctuations, a feature that is not captured by the Bogoliubov approximation (see Eq.~(\ref{eq:dragF})).

\section{Measuring the drag via a quick perturbation}\label{sec:quickresponse}
Collective excitations induced by a static perturbation  (where $\omega=0$) are governed both by the drag, through the moment $M_{1,s}$ or $M_{1,D_s}$, and the susceptibility of the system, through the moment $M_{-1,s}$ or $M_{-1,D_s}$.
As we showed in Fig. (\ref{fig:speedofsound}) and Fig. (\ref{fig:spindipole}) the two effects are of comparable magnitude, so it is convenient to devise an experiment where only the drag effect plays a role.
In addition, while it would be desirable to study the effect in strongly interacting mixtures, three body losses will significantly abbreviate the stability of such systems.

In order to overcome both these limitations we consider the linear response of a trapped gas to a perturbation that is suddenly turned on.

We start by considering the the frequency response $\chi_{D_A,D_B}(\omega)$ of the operator ${\hat{D}_B=\int d^3 \mathbf{r} \, x \,\hat{n}_B(\mathbf{r})}$ to a perturbation ${\hat{D}_A=\int d^3 \mathbf{r} \, x \,\hat{n}_A(\mathbf{r})}$, which is given by:
\begin{equation}
\begin{split}
\chi_{D_A,D_B}(\omega) = & \\ \frac{2}{V} 
\sum_{n\neq 0} & \frac{\braket{ n|\hat{D}_A|0 } \braket{ 0|\hat{D}_B|n }}{\omega^2+(E_0-E_n)^2
} \left(E_0-E_n\right) \, .
\end{split}
\end{equation}
The response function satisfies the following expansion at high frequencies \cite{StrPit}:
\begin{equation}
\begin{split}
\chi_{D_A,D_B}(\omega) & =  -\frac{1}{\omega^2 V}\braket{\left[\hat{D}_B, \left[\hat{H}_{eff},\hat{D}_A\right] \right]}+O\left( \frac{1}{\omega^3} \right)  = \\ &
= - \frac{1}{\omega^2 V} \int d^3 \mathbf{r}\, \frac{\rho_{AB}(\mathbf{r})}{m^2} +O\left( \frac{1}{\omega^3} \right) \, ,
\end{split}
\end{equation}
where in the last equality we made use of Eq.~(\ref{eq:m1trap}).
We thus see that the response function at high frequencies is directly proportional to the drag.
This describes a situation where the probe $\hat{D}_A$ is quickly turned on and the observable $\hat{D}_B$ is measured shortly after, as we will clarify in what follows.

Consider a time dependent perturbation of the form:
\begin{equation}
\hat{V}(t)= \lambda \hat{D}_A \, \theta(t) \, ,
\end{equation}
where $\theta(t)$ is the Heaviside step function and $\lambda$ is a small parameter.
The variation of the average of $D_B$ to first order in $\lambda$ at time $t$ is given by:
\begin{equation}
\braket{\delta D_B}(t)= \lambda V \int_{-\infty}^\infty dt' \chi_{D_A,D_B} (t-t') \theta(t') \, .
\end{equation}
Using the Fourier transform of the theta function $\tilde{\theta}(\omega)= \frac{1}{i\omega} + \pi \delta (\omega)$ and after some manipulation we arrive at
\begin{equation}\label{eq:expand}
\braket{\delta D_B}(t)= -\frac{\lambda V}{2 \pi} \int d\omega \, \frac{e^{-i\omega t}-1}{i\omega} \chi_{D_A,D_B} (\omega) \, ,
\end{equation}
where we made use of the useful identities  ${\pi \chi_{D_A,D_B} (0)=i\int d\omega \, \chi_{D_A,D_B}(\omega)/\omega}$ and ${\int d\omega \, \chi_{D_A,D_B} (\omega)=0}$.
While this integral is in principle over all frequencies $\omega$, the response function $\chi_{D_A,D_B} (\omega)$ will have a cutoff in frequency $\overline{\omega}$ above which it vanishes.
We thus expand this expression for short times, i.e. times $t$ such that ${t \ll 1/\overline{\omega}}$:
\begin{equation} \label{eq:quickres}
\braket{\delta D_B}(t)= -i\frac{\lambda V}{4 \pi} \int_{-\infty}^\infty d\omega \, \omega t^2 \chi_{D_A,D_B} (\omega) \, .
\end{equation}

Since the real part of the response function is an even function of $\omega$ we can express the response in terms of the imaginary part of the response function $\chi''_{D_A,D_B} (\omega)$ which at zero temperature satisfies the sum rule \cite{StrPit}:
\begin{equation}
\int_{-\infty}^\infty d\omega \, \omega \, \chi''_{D_A,D_B} (\omega)=\frac{\pi}{V} \braket{\left[\hat{D}_A, \left[\hat{H}_{eff},\hat{D}_B\right] \right]} \, .
\end{equation}
Using this equality in Eq.~(\ref{eq:quickres}) and making use of Eq.~(\ref{eq:m1trap}) we finally obtain the desired equation in terms of the double commutator:
\begin{equation}
\begin{split}
\braket{\delta D_B}(t) & = \frac{\lambda}{4}  t^2  \braket{\left[\hat{D}_A, \left[\hat{H}_{eff},\hat{D}_B\right] \right]}= \\& 
=\frac{ \lambda t^2}{ 4} \int d^3 \mathbf{r}\, \frac{\rho_{AB}(\mathbf{r})}{m^2} \, .
\end{split}
\end{equation}

This calculation outlines a useful experimental procedure to measure the drag coefficient. A superfluid mixture subject to a dipole moment $\hat{D}_A$ that is suddenly turned on will have the other component increase ballistically its dipole moment $\hat{D}_B$ for short times, with a coefficient proportional to the drag coefficient $\rho_{AB}$.
This experimental procedure would allow to measure the drag directly, without the need to independently measure the susceptibility of the system (as in secs.~\ref{sec:sound} and \ref{sec:dipole}).
Importantly, the short time scales involved significantly reduce the amount of three body losses.
This makes the procedure applicable to strongly interacting systems where the effect is more sizeable.

\section{Conclusions}
In this work we analysed the Andreev-Bashkin effect in a two species superfluid within linear response theory.
In analogy with the single component case the superfluid densities can be expressed in terms of transverse current-current response functions.
A striking result that has no analogy with the single component case is that, while the overall response of the superfluid to a transverse vector field vanishes at zero temperature, such a field will give rise nonzero response even at vanishing temperature when acting only on one component.
In this sense, the drag behaves as a sort of normal component, inducing a current response in reaction to a transverse probe.

The presence of a finite drag density arises from multiparticle excited states, which are shown to give an additional contribution to the first moment of the spin structure factor $S_s(\mathbf{q},\omega)$.
This result is also in contrast with the case of single component superfluids, where single particle excitations are the only low-lying excited states at zero temperature.
Since the moments of $S_s(\mathbf{q},\omega)$ are constrained by a sum rule, the fact that the drag is nonzero means that the single quasiparticle contribution is changed (see e.g. Eq.~(\ref{eq:dragsumrule})).
Using current-current response functions we evaluate the collisionless drag in a two component weakly interacting Bose gas within the Bogoliubov approximation. This allows us to easily recover the results obtained via 
energy vacuum calculations \cite{FilShe} and give a more direct interpretation of the drag as spin-density phonon mixing. 
We show how typical measurable quantities as the spin speed of sound and the spin dipole mode frequency  \cite{KimShi,Ferrari2018} are affected by the presence of the drag. While the change in the susceptibility 
due to quantum fluctuations dominates the correction to the spin speed of sound, in the case of the spin dipole frequency the correction due to the presence of the drag can be of the same order of magnitude as the correction coming from the susceptibility.

We show that the drag can be directly measured in an experiment where a dipole moment is quickly induced on one component. This induces at short times a dipole moment in the other component which is proportional to the drag density.

Our analysis can be replicated to compute the drag and its effect on observables in strongly interacting systems.
This is particularly relevant for new cold atoms systems as Bose-Bose mixtures on optical lattices \cite{FirenzeBB, Schneble, Bloch}, superfluid Bose-Fermi mixtures \cite{SalomonBF} and in the foreseeable future superfluid Fermi-Fermi mixtures
and the short-lived strongly interacting Bose mixtures (as an extension of the recently realized strongly interacting Bose gas \cite{Hadzibabic,Ching}).

\section*{Acknowledgements}
This project was supported by the University of Southampton as host of the Vice-Chancellor Fellowship scheme, by the Provincia Autonoma di Trento, the Fis$\hbar$ project of the Istituto Nazionale
di Fisica Nucleare. 
We thank S. Stringari and G. D'Alessandro for discussions.

\appendix \label{appendix}
\section{Thermodynamic quantities for the weakly interacting Bose-Bose mixture}\label{app:thermodynamics}
In this appendix we derive the quantities that are needed to derive the inverse energy weighted sum rule used in subsecs.~\ref{sec:sound} and \ref{sec:dipole}. 

We start by considering a homogeneous weakly interacting Bose-Bose mixture as in subsec.~\ref{sec:sound}.
The energy of a Bose-Bose mixture in the mean field approximation is just given by:
\begin{equation}\label{eq:mfenergy}
\frac{E_{MF}}{V}=\frac{1}{2}g_{AA}n_A^2+\frac{1}{2}g_{BB}n_B^2+g_{AB} n_A n_B \, .
\end{equation}
The lowest order beyond mean field correction to this energy, known as the Lee-Huang-Yang term, can be found in \cite{Petrov} and reads for equal masses $m_A=m_B=m$ and intraspecies coupling $g_{AA}=g_{BB}=g$:
\begin{equation}\label{eq:LHY}
E_{LHY}/V=\frac{8}{15\pi^2} m^{3/2} \left(g n_A \right)^{5/2} f\left(1, \frac{g_{AB}^2}{g^2}, \frac{n_B}{n_A}\right) \, ,
\end{equation}
with $f(1,x,y)=\sum_{\pm}\left( 1+y \pm \sqrt{(1-y)^2+4xy} \right)^{5/2}/4\sqrt{2}$,
a dimensionless function.
%The function $f$ can be recast into a function of the polarization $p=\frac{n_{A}-n_{B}}{n_{A}+n_{B}}$ with the substitution $y=\frac{1-p}{1+p}$.

From equation (\ref{eq:LHY}) we can obtain the chemical potential $\mu_\alpha=\frac{\partial E}{\partial N_{\alpha}}$ of each species $\alpha$.
In the $\mathbb{Z}_2$ symmetric case where $\mu_A=\mu_B=\mu$ we get:
\begin{equation}
\mu=\left(g+g_{AB}\right)n \left(1+\frac{2}{3}\sqrt{na^3} B(\eta))\right) \, ,
\end{equation}
where $\eta=\frac{|g_{AB}|}{g}$ and:
\begin{equation}\label{eq:B}
B(\eta)=\frac{8}{\sqrt{\pi}(1+\eta)}\left((1+\eta)^{5/2}+(1-\eta)^{5/2}\right) \, .
\end{equation}
We can also derive the magnetic susceptibility ${\chi_s^{-1}= \frac{\partial^2(E/V)}{\partial(n_A-n_B)^2}}$, which reads:
\begin{equation}\label{eq:susceptibility}
\chi^{-1}_s  =  g\frac{1-\eta}{2}\left[1+ \sqrt{na^3}  C(\eta) \right] \, ,
\end{equation}
where we defined the function
\begin{equation}\label{eq:C}
C(\eta)=\frac{16}{3\sqrt{\pi}} \frac{1+\eta}{\eta}\left( (1+\eta)^{3/2} - (1-\eta)^{3/2}) \right) \, .
\end{equation}

For the case of the spin dipole oscillations of subsec.~\ref{sec:dipole} we need instead to consider the trapped case.
In presence of harmonic trapping and within the local density approximation the chemical potential $\mu$ and the susceptibility become position-dependent through the equilibrium density profile $n(\mathbf{r})$. At mean field level the latter reads:
\begin{equation}
n(\mathbf{r})=\frac{2\mu_{TF}}{\left( g+g_{AB}\right)}\left(1-\frac{r^2}{R_{TF}^2}\right) \, ,
\end{equation}
where $R_{TF}=\sqrt{\frac{2 \mu_{TF}}{m \omega_0^2}}$ is the Thomas-Fermi radius and $\mu_{TF}$ the Thomas-Fermi chemical potential.
Imposing that the integrated density gives the total number of particles we find the Thomas-Fermi radius:
\begin{equation} \label{eq:thomasfermi2nd}
R_{TF}=\left(\frac{15N \left(g+g_{AB} \right)}{8 \pi m \omega_0^2}\right)^{1/5} \, ,
\end{equation}
from which we can derive the chemical potential as a function of the total number of particles, the trapping potential and the interactions.

The inclusion of beyond mean fields effects results in a correction in the value of $\mu_{TF}$ and of the Thomas-Fermi radius $R_{TF}$.
The chemical potential to first order in the gas parameter $\sqrt{n(0)a^3}$ reads:
\begin{equation}
\mu_{TF}^1=\mu_{TF}\left(1+\frac{\pi}{16} \sqrt{n(0)a^3} B(\eta) \right) \, ,
\end{equation}
and as a consequence:
\begin{equation}\label{eq:thomasfermi2}
R^1_{TF}=R_{TF}\left(1+\frac{\pi}{32} \sqrt{n(0)a^3} B(\eta) \right) \, .
\end{equation}
The result for the Thomas-Fermi radius is used to compute the integral in Eq.~(\ref{eq:m-1trap}).

\section{2-component Bose-Hubbard model}\label{subsec:latticedrag}

In this appendix we derive the drag in a 2-component Bose-Hubbard model, recovering a result contained in \cite{LinSub}.
In particular we derive the equations for the superfluid and normal stiffnesses in a system lacking Galilean invariance, where one loses the normalization condition of Eq.~(\ref{eq:normalization2}).
The drag in the weakly interacting case is not enhanced in the Bose-Hubbard model compared to the continuum case~\cite{LinSub}. 
The introduction of the lattice can be interesting to reach regimes where the correlations are strong but the condensate is still stable with respect to three-body losses~\cite{Bloch}.

We consider a weakly interacting Bose-Bose mixture of $N$ bosons in a cubic lattice of volume $V$ with periodic boundary conditions, in the thermodynamic limit. We call $l$ the lattice spacing and $I$ the number of sites.
The tunneling coefficient $J$ is the same for both species.
The Bose-Hubbard Hamiltonian describing the system is:
\begin{equation}\hat{H}=\hat{H}_A+\hat{H}_B+\hat{H}_{AB}  \, ,
\end{equation}
with
\begin{equation}
\hat{H}_A= -J\sum_{\left< i,j \right>} ( \hat{a}_{i,A}^{\dagger} \hat{a}_{j,A}+h.c.)+U\sum_{i} \hat{n}_{i,A}(\hat{n}_{i,A}-1) \, ,
\end{equation}
\begin{equation}
\hat{H}_B =  -J\sum_{\left< i,j \right>} ( \hat{a}_{i,B}^{\dagger} \hat{a}_{j,B}+h.c.)+U\sum_{i} \hat{n}_{i,B}(\hat{n}_{i,B}-1) \, ,
\end{equation}
\begin{equation}
\hat{H}_{AB}=U_{AB}\sum_{i} \hat{n}_{i,A}\hat{n}_{i,B} \, ,
\end{equation}
where $\hat{a}_{i,\alpha}^{\dagger}$ and $\hat{a}_{i,\alpha}$ are bosonic creation and annihilation operators at each site $i$ and $\hat{n}_{i,\alpha} =\hat{a}_{i,\alpha}^{\dagger}\hat{a}_{i,\alpha}$ is the number of particles of species $\alpha$ at site $i$.
From the tunneling coefficient $J$ we can define the usual effective mass as $m^*=\hbar^2/(2Jl^2)$.
The current density operator at a given site $\mathbf{r}$ on the cubic lattice is defined as:
\begin{equation}
\hat{\mathbf{j}}_{\alpha}(\mathbf{r})=-i\frac{J l}{ \hbar V} \sum_{\mathbf{u}}( \hat{a}_{\mathbf{r},\alpha}^{\dagger} \hat{a}_{\mathbf{r}+\mathbf{u},\alpha}-h.c.)\mathbf{u} \, ,
\end{equation}
where $\mathbf{u}$ is a lattice vector of unit length.
The linear response treatment of the three-fluid hydrodynamics in the case of a lattice is very similar to that of subsec.~\ref{subsec:currentcurrent}.
The difference is in the transformation law of the current operator and of the Hamiltonian in presence of a vector potential $\mathbf{A}(\mathbf{r})$.
The transformation for the current, corresponding to Eq.~(\ref{eq:transformationj}) in the continuous model:
\begin{equation}\label{eq:transformationjdiscrete}
\hat{\mathbf{j}}_\alpha(\mathbf{r}) \rightarrow \hat{\mathbf{j}}_\alpha(\mathbf{r}) - \sum_{\mathbf{u}} \frac{e_\alpha}{m_\alpha} \frac{\hat{K}_{u,\alpha}(\mathbf{r})}{2J} A_u(\mathbf{r}) \, \mathbf{u} \, ,
\end{equation}
where $\hat{K}_{u,\alpha}(\mathbf{r})$ is the kinetic energy density operator for a species $\alpha$ along $\mathbf{u}$ directed links, namely
\begin{equation}
\hat{K}_{u,\alpha}(\mathbf{r})=- \frac{J}{V} ( \hat{a}_{\mathbf{r},\alpha}^{\dagger} \hat{a}_{\mathbf{r}+\mathbf{u},\alpha}+h.c.) \, .
\end{equation}
For a system with discrete translational invariance we have that $\left< \hat{K}_{u,\alpha}(\mathbf{r})\right>=K_{u,\alpha}$.
The Hamiltonian gets transformed to (compare with Eq.~(\ref{eq:transformationH})):
\begin{equation}\label{eq:transformationHdiscrete}
\begin{split}
\hat{H}\rightarrow  \hat{H} & -\sum_{\alpha=A,B} \sum_{\mathbf{r}} \frac{e_\alpha}{m_\alpha} \hat{\mathbf{j}}_{\alpha}(\mathbf{r}) \cdot  \mathbf{A}(\mathbf{r}) +\\ & 
+\sum_{\alpha=A,B}\sum_{\mathbf{r},u} \frac{e_\alpha^2}{2m_\alpha}  \frac{\hat{K}_{u,\alpha}(\mathbf{r})}{2J}  A^2(\mathbf{r}) \, .
\end{split}
\end{equation}

Applying linear response theory as in subsec.~\ref{subsec:currentcurrent} we obtain:
\begin{equation}\label{rhodrag1lattice}
-m^{*2} \lim_{\mathbf{q} \to 0} \chi_{\mathbf{j}_{A},\mathbf{j}_{B}}^{T}(\mathbf{q})=   \rho_{AB} \, ,
\end{equation}
\begin{equation}\label{rhodrag2lattice}
m^*\left(-\frac{1}{2J}\braket{\hat{K}_{u,\alpha}}-m^*\lim_{\mathbf{q} \to 0}\chi_{\mathbf{j}_{\alpha},\mathbf{j}_{\alpha}}^{T}(\mathbf{q})\right)=   \rho_{\alpha\alpha} \, ,
\end{equation}
\begin{equation}\label{rhodrag3lattice}
 m^{*2}\sum_{\alpha,\beta=A,B} \lim_{\mathbf{q} \to 0} \chi_{\mathbf{j}_{\alpha},\mathbf{j}_{\beta}}^{T}(\mathbf{q})=   \rho_n \, .
\end{equation}
Eqs.~(\ref{rhodrag1lattice}), (\ref{rhodrag2lattice}) and (\ref{rhodrag3lattice}) are the discrete space analogous to Eqs.~(\ref{rhodrag1}), (\ref{rhodrag2}) and (\ref{rhodrag3}) of sec.~\ref{sec2}.
The only difference is the operator $ \hat{K}_\alpha/2J$ in place of the density $\hat{n}_\alpha$, as a result of the different transformation rule of the current $\hat{\mathbf{j}}(\mathbf{r})$ and the Hamiltonian $\hat{H}$ under a vector potential in presence of a lattice, Eqs.~(\ref{eq:transformationjdiscrete}) and (\ref{eq:transformationHdiscrete}).

We can compute the drag in the Bogoliubov approximation as we did in  the translational invariant system to obtain Eq.~(\ref{final}):
\begin{equation}\label{latticedrag}
\begin{split}
\rho_{A B}
=\frac{\hbar^2}{8V}\sum_{\mathbf{k}} \frac{(\Omega_{d,\mathbf{k}}-\Omega_{s,\mathbf{k}})^2 k_x^2}{(\Omega_{d,\mathbf{k}}+\Omega_{s,\mathbf{k}})\Omega_{s,\mathbf{k}}\Omega_{d,\mathbf{k}}}\left(\frac{\sin(k_x l)}{k_x l}\right)^2 \, ,
\end{split}
\end{equation}
where $\Omega_{d,\mathbf{k}}$ and $\Omega_{s,\mathbf{k}}$ are the excitation energies of the density and spin degrees of freedom respectively, namely:
\begin{equation}
\Omega_{d (s),\mathbf{k}}=\sqrt{\epsilon(\mathbf{k})(\epsilon(\mathbf{k})+2 U f\pm 2U_{AB} f)} \, ,
\end{equation}
with $\epsilon(\mathbf{k})=4J\sum_{i=x,y,z}\sin^2(k_il /2)$ and $f=N/2I$ is the filling fraction for each species. In the limit $l \to 0$, Eq.~(\ref{latticedrag}) coincides with the result for the homogeneous system, Eq.~(\ref{final}). 

In \cite{Contessi} the Bogoliubov result for the drag has been compared with a microscopic calculation for a  one-dimensional system. Interestingly Eq.~(\ref{latticedrag})
turns out to give a reasonable estimate of the drag even in this case provided $U_{AB}>0$.
\bibliography{bibliografia}
\end{document}